\documentclass[12pt]{iopart}
\usepackage{graphicx}

\begin{document}

\title{Percolation of overlapping squares or cubes on a lattice}
\author{Zbigniew Koza, Grzegorz Kondrat, and Karol Suszczy\'nski}
\address{Faculty of Physics and Astronomy, University of Wroc{\l}aw, 50-204 Wroc{\l}aw,
Poland}
\ead{zbigniew.koza@ift.uni.wroc.pl}

\date{\today}

\begin{abstract}
Porous media are often modelled as systems of overlapping obstacles,
which leads to the problem of two percolation thresholds in such systems,
one for the porous matrix and the other one for the void space. Here we investigate
these percolation thresholds in the model
of overlapping squares or cubes of linear size $k>1$ randomly distributed on a regular lattice.
We find that the percolation threshold of obstacles is a nonmonotonic function of $k$,
whereas the percolation threshold of the void space is well
approximated by a function linear in $1/k$.
We propose a generalization of the excluded volume approximation to discrete systems and use it to investigate
the transition between continuous and discrete percolation,
finding a remarkable agreement between the theory and numerical results.
We argue that the continuous percolation threshold of aligned squares on a plane
is the same for the solid and void phases and estimate the continuous percolation threshold of the void space around aligned cubes
in a 3D space as  0.036(1). We also discuss the connection of the model to the standard site percolation with complex neighborhood.

\end{abstract}


\submitto{\JSTAT}


\maketitle

\section{Introduction\label{sec:intro}}
One of the standard ways of constructing a nontrivial model of a  heterogeneous
porous material is to fill the space randomly with a number of identical impermeable
objects that are free to overlap \cite{Torquato2002}.
Models of this type
can be used to derive exact bounds on the values of
various transport coefficient, including the effective diffusion coefficient,
electrical conductivity, dielectric constant and  magnetic permeability of dispersions \cite{Weissberg63,Maxwell},
whereas at high obstacle concentrations they can be used to model
consolidated porous media, e.g., sandstones, sintered materials, rocks, or complex catalysts
\cite{Torquato2002,Voutilainen2013,Zalc2003,Mace1991}.

Systems of identical overlapping objects
are often described with the reduced number density
$
  \eta = vN/V,
$
where $v$ is the area (or volume) of each object, $N$ is their number, and
$V$ is the system volume \cite{Baker2002}.
As $\eta$ is gradually increased from 0,
the system undergoes two critical transformations. The first one is
related to the connectivity of the solid phase formed by the obstacles:
at low densities the obstacles form a disconnected phase, whereas
above a critical value  $\eta_\mathrm{s}$ this phase becomes connected.
The second one is related to the connectivity of the void space: above some $\eta_\mathrm{v} \ge \eta_\mathrm{s}$
the void space becomes disconnected and the transport is blocked.
Clearly, $\eta_\mathrm{s}$ and $\eta_\mathrm{v}$ can be identified as  the percolation thresholds
for the solid and the void phase, respectively.

So far a vast majority of research has focused on the simplest case of spherical obstacles.
Spheres (in 3D) and disks (in 2D) are particularly convenient in the mathematical treatment of
problems where the transport is governed by the Laplace's equation, e.g., diffusion \cite{Weissberg63}.
The task of determining $\eta_\mathrm{s}$ is then equivalent to the well known
problem of finding the percolation threshold for overlapping spheres  \cite{Torquato2012a,Torquato2012b}
or disks \cite{Torquato2012a,Torquato2012b,Garboczi1991,Asikainen2000}. Similarly, the value of $\eta_\mathrm{v}$ can be identified as
the percolation threshold in the ``Swiss-cheese'' model for spheres \cite{Lorenz2001,Priour2014,Marck1996} or disks
\cite{Marck1996,Okazaki1996}, in which percolation is sought for in the void phase.

However, other shapes are also important \cite{Balberg1987}, e.g., needles \cite{Garboczi1991,Asikainen2000,Provatas2000,Mertens2012},
fibers \cite{Provatas2000},
ellipses \cite{Xia1988}, and ellipsoids \cite{Yi2006}.
In particular, many computer simulations
discretize the space into a regular lattice and model a porous medium using
overlapping cubes \cite{Kohout2004,Matyka2013} or
squares \cite{Koponen1996,Koponen1997,Matyka2008,Jiang2008}.
Therefore we investigate the problem of the percolation thresholds
in the model of overlapping squares (2D) or cubes (3D) on a regular lattice,
focusing on their dependence on the size of the obstacles.
While one might expect that the values of both percolation thresholds
for finite
values of the obstacle size $k$ could be approximated by interpolating their values between the
well known cases of $k=1$ (standard site percolation on a lattice)  and $k\to\infty$
(off-lattice percolation of freely overlapping squares or cubes), we found that this is not the case
and the percolation threshold of the obstacles exhibits a maximum at a finite value of $k>1$.
This result is particularly important for designing numerical models of porous media close to percolation.

By investigating the limit of $k\to\infty$, the model studied here can be used to
explore the transition between discrete and continuous percolation.
On the other hand, the dependency of continuous percolation on the obstacle shape
can be investigated with a simple theory, the excluded volume approximation \cite{Balberg1984,Balberg1987}.
We combine the two facts and generalize the excluded volume theory to discrete systems,
finding a remarkably good agreement with numerical results for all obstacle sizes $k$.
By extrapolating the numerical results obtained for discrete systems, we
improve the accuracy of the determination of the continuous percolation threshold
of aligned, overlapping cubes in three dimensions as well as provide the values of continuous percolation thresholds
of the void space around aligned squares and cubes in dimensions two and three.
Moreover, we show that a direct relation  between the model of overlapping
objects of arbitrary shape and the standard site percolation
with complex neighborhood can be used to improve the accuracy of the percolation
threshold determination in a class of percolation models.


\section{The Model\label{sec:Model}}

\begin{figure}
 \begin{center}
  \includegraphics[width=0.9\columnwidth]{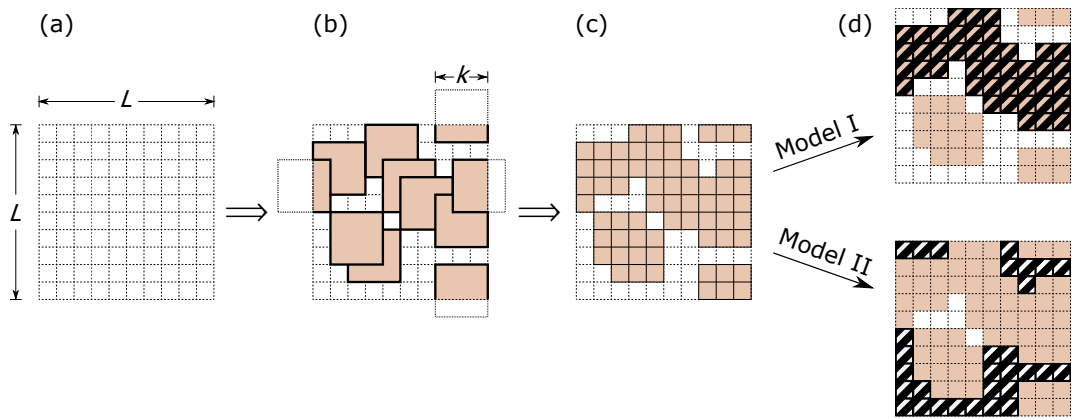}   
  \caption{ \label{fig:x}
            Construction of the models.
            An empty regular lattice of size $L\times L$ lattice units (l.u.) with periodic boundary conditions (a)
            is filled at random with  square obstacles of size $k\times k$~l.u.\
            aligned to the lattice axes (b) and the elementary cells occupied by the obstacles are identified (c); finally,
            the problem is reduced to the site percolation through the occupied or unoccupied
            elementary cells in Model I and Model~II, respectively (d).
            The same method was also used in three dimensions.
  }
 \end{center}
\end{figure}

As illustrated in figure~\ref{fig:x},
we consider a square (2D) or cubic (3D) lattice of linear size $L$ lattice units,
on which square or cubic ``obstacles'' of linear size $k$ lattice units
are deposited at random,
where $k,L$ are some
integer parameters, $1\le k\ll L$,
and the periodic boundary conditions are imposed on the system
during the obstacle deposition to reduce the finite-size effects \cite{Stauffer1994,Isichenko1992}.
The obstacles are aligned to the underlying lattice, their edges coinciding
with  lattice nodes, and are free to overlap.
The volume occupied by obstacles is thus a simple union of elementary lattice cells
and the model is essentially discrete.
Henceforth the volume occupied by obstacles shall be called phase A or solid phase,
whereas the remaining volume shall be called phase B or void phase.

We investigate two cases, which will be denoted as Model I and Model II. In Model I
we consider the percolation of obstacles, whereas in Model II
we consider the percolation of the void space. Two elementary cells are considered to be connected
directly if and only if they belong to the same phase and share the same side (2D) or face (3D).
We define a percolation cluster as a set of elementary cells spanning two opposite system sides
through a sequence of directly connected elementary cells.
Thus, Model I interpolates between the standard model of site percolation on a regular lattice  ($k=1$)
and the model of continuous percolation of aligned squares \cite{Baker2002,Mertens2012}
or cubes \cite{Baker2002} ($k\to\infty$).
Similarly, Model II interpolates between the site percolation of voids on a regular lattice $(k=1)$ and the
continuous percolation of voids around aligned squares  or cubes   ($k\to\infty$).

We shall express the percolation threshold in terms of the porosity, $\varphi$, which is one  of
the most fundamental concepts in porous media studies.
In both Model I and II the porosity is defined as the ratio of the volume occupied
by phase B (void space) to the system volume.
For Model~II this quantity is equivalent to the so called remaining area/volume
fraction \cite{Xia1988}, whereas  $\varphi = 1 - p_\mathrm{A}$
for model~I, with $p_\mathrm{A}$
being the area (2D) or volume (3D) fraction of phase A.

Since locations of each obstacle are random and mutually independent, and since the porosity can be
identified with the probability that an elementary cell is unoccupied by any obstacle,
the expected porosity, $\langle \varphi \rangle$, can be expressed as
\begin{equation}
  \label{eq:<phi>}
  \langle \varphi \rangle = (1 - v/V)^N = (1-\eta/N)^N.
\end{equation}
This formula can be used to generate a system of a given porosity;
it also implies that
\begin{equation}
  \label{eq:eta}
      \varphi = \exp(-\eta)
\end{equation}
at the thermodynamic limit \cite{Baker2002}.


\section{Numerical and mathematical details}

To find the percolation thresholds we used two independent computer programs.
One of them keeps track of whether each elementary cell belongs to phase A or B.
In this approach the execution time of each simulation is $\propto L^d$, where $d$ is the space dimensionality,
 and we were able to run it for
$L\le 4096$ ($d=2$) and $L\le 1024$ ($d=3$). We used this program to simulate Model II
(percolation of voids).

The second program uses data structures typical of the algorithms designed for the continuous percolation:
each obstacle is identified by its coordinates, i.e., by only $d$ integers, and the clusters are found using the union-find algorithm.
In this case the computer memory storage, as well as the simulation time of each percolation cluster,
is $\propto (L/k)^d$ and we were able to run the program  for
$L/k\le 5000$ ($d=2$) and $L/k\le400$ ($d=3$). While the second program allows for much larger
system sizes $L$, it cannot be efficiently used for Model II.
We used it to simulate Model I (percolation of obstacles).

\begin{figure}
 \begin{center}
  \includegraphics[width=0.9\columnwidth]{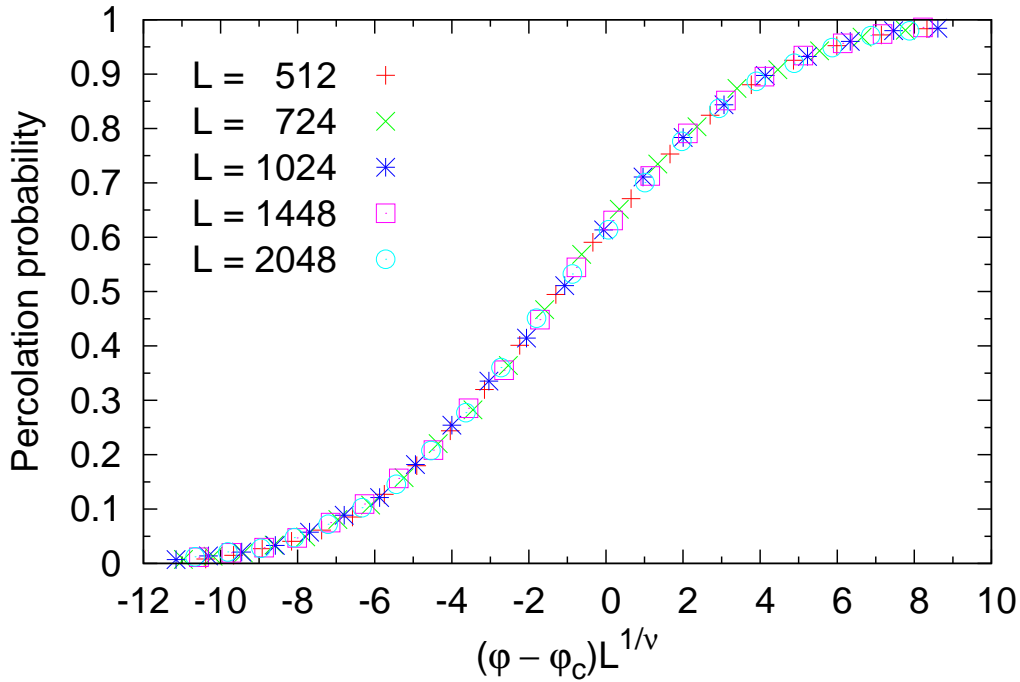}   
  \caption{ \label{fig:1}
            The data collapse according to the universal scaling, equation (\protect\ref{eq:scaling})
            with $\varphi^\mathrm{c}_k = 0.3505$, $\nu = 4/3$,
            obtained for Model II with the obstacle size $k=20$ and space dimension $d=2$.}
 \end{center}
\end{figure}

In each case we assumed periodic boundary conditions in the direction(s) perpendicular
to the direction along which we sought the percolation. The number of independent
samples ($N$) varied from $10^7$ for very small systems to $10^3$ for larger $L$ so as to ensure that
we can determine the percolation threshold for given values of $L$ and $k$ with the absolute accuracy
of at least $5\times 10^{-5}$. To this end we determined the probability $P_{L,k}(\varphi)$
that a system of size $L$, obstacle size $k$, and porosity $\varphi$ contains a percolating (spanning) cluster.
In accordance with the finite-size scaling theory,
$P_{L,k}$ is expected to scale with $L$ and the deviation $\varphi - \varphi^\textrm{c}_k$
from  the critical porosity $\varphi^\textrm{c}_k$ as \cite{Stauffer1994,Cornette2003}
\begin{equation}
  \label{eq:scaling}
  P_{L,k}(\varphi) = f_k\left(
                         (\varphi-\varphi^\textrm{c}_k)L^{1/\nu}
                \right), \quad  L/k \gg 1,
\end{equation}
where $\nu$ is the correlation length exponent and $f_k$ is a scaling function,  cf.\ figure~\ref{fig:1}.
Then we define an effective, $L$-dependent critical porosity  $\varphi^\textrm{c}_k(L)$
as the solution to $P_{L,k}(\varphi) = 0.5$. This value can be obtained, for example, by fitting the data
to a curve with a similar, sigmoidal shape, e.g.,
\begin{equation}
  \label{eq:PLk}
    P_{L,k}(\varphi) \approx \frac{\mathop{\mathrm{erfc}}\{[\varphi^\textrm{c}_k(L)- \varphi]/\Delta(L)\}}{2}.
\end{equation}
Not only does this method provide an accurate estimation of   $\varphi^\textrm{c}_k(L)$, but
it also yields the width  of the percolation transition, $\Delta(L)$,
which can be used to estimate the value of the critical exponent $\nu$ from
\begin{equation}
  \label{eq:Delta-scaling}
  \Delta(L) \propto L^{-1/\nu}, \quad L \gg 1.
\end{equation}
Once $\varphi^\textrm{c}_k(L)$ have been determined for sufficiently large values of $L$,
they can be used to obtain the critical porosity
$\varphi^\textrm{c}_k$ from
\begin{equation}
 \label{eq:phi-scaling}
 \varphi^\textrm{c}_k(L) - \varphi^\textrm{c}_k \propto L^{-1/\nu}, \quad L \gg 1.
\end{equation}


\section{Results\label{Sec:Results}}

\subsection{Asymptotic regime and exponent $\nu$ \label{sub:asymptotic}}
The critical porosity can be determined from (\ref{eq:phi-scaling}) by
fitting $\varphi^\textrm{c}_k(L)$ to a nonlinear function $\varphi^\textrm{c}_k + a L^{-1/\nu}$ with three
fitting parameters, $\varphi^\textrm{c}_k$, $a$ and $\nu$. The fourth unknown parameter, $L_\mathrm{min}$
is the minimum value of $L$ for which the asymptotic regime expressed by (\ref{eq:phi-scaling}) holds with a
sufficient accuracy.  However, the value of the critical exponent $\nu$ is expected to be universal, $k$-independent.
We verified this hypothesis positively for Models I and II in  dimensions $2$ and $3$ for all tested values of $k$
using both (\ref{eq:Delta-scaling}) and (\ref{eq:phi-scaling}) (data not shown).
Therefore in the fitting procedure we reduced the number of unknowns to two, assuming the most accurate value of
the thermal exponent $y_\mathrm{t} = {1/\nu}$ available for the standard site percolation,
$1/\nu = 3/4$ and $1.141$ for $d=2$ and $3$, respectively \cite{Xu2014,Wang2013}.
The value of $L_\mathrm{min}$ was determined by requiring that the reduced chi-squared statistic of the fit is $\sim O(1)$
and neither this parameter nor the estimated value of the critical porosity
varies significantly for $L > L_\mathrm{min}$.  For Model I we found $L_\mathrm{min} = 100k$ and $60k$
for $d=2$ and $3$, respectively, whereas for Model II we found  $L_\mathrm{min} = 50k$
for $d=2,3$. The accuracy of our data was insufficient
for application of fits with higher-order terms,
a standard technique for simpler models of percolation \cite{Xu2014,Wang2013}.

\begin{figure}
 \begin{center}
  \includegraphics[width=0.95\columnwidth]{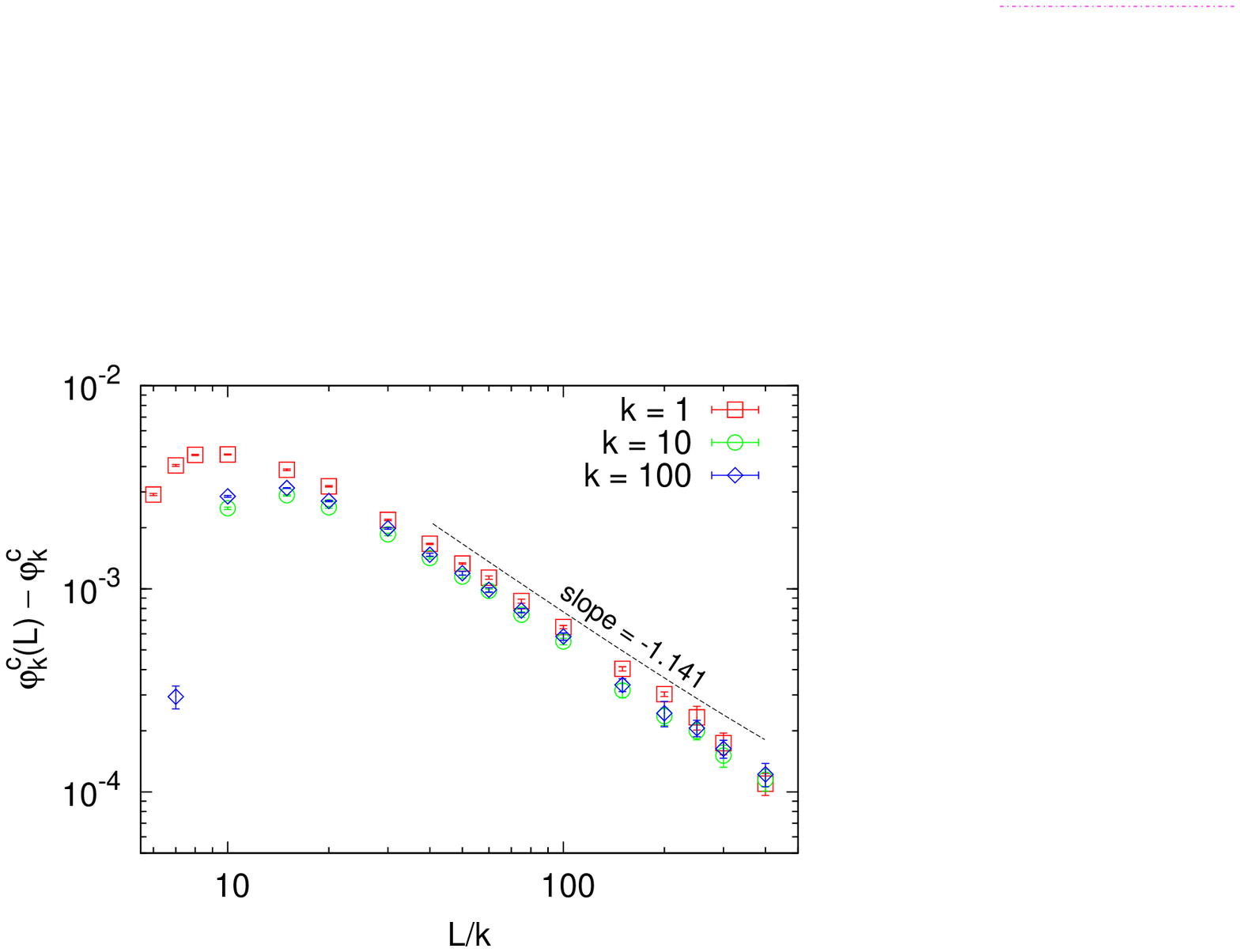}  
  \caption{ \label{fig:convergence}
     (Color online) Convergence of the effective critical porosity $\varphi^\textrm{c}_k(L)$
     to its asymptotic value  $\varphi^\textrm{c}_k$ for Model I in three dimensions and $k=1,10,100$.
     The dashed line is a guide to the eye with the slope $-1/\nu = -1.141$ determined
     for the standard site percolation ($k=1$, $d=3$) in Ref.~\protect\cite{Xu2014,Wang2013}.}
 \end{center}
\end{figure}

Selected results showing the convergence of the effective critical porosity to its asymptotic value
are depicted in figure~\ref{fig:convergence}.
As can be seen, the convergence rate is similar for different values of $k$.


\subsection{Critical porosity\label{sec:porosity}}

\subsubsection{Model I \label{sub:Model-I}}

\begin{figure}
 \begin{center}
  \includegraphics[width=0.95\columnwidth]{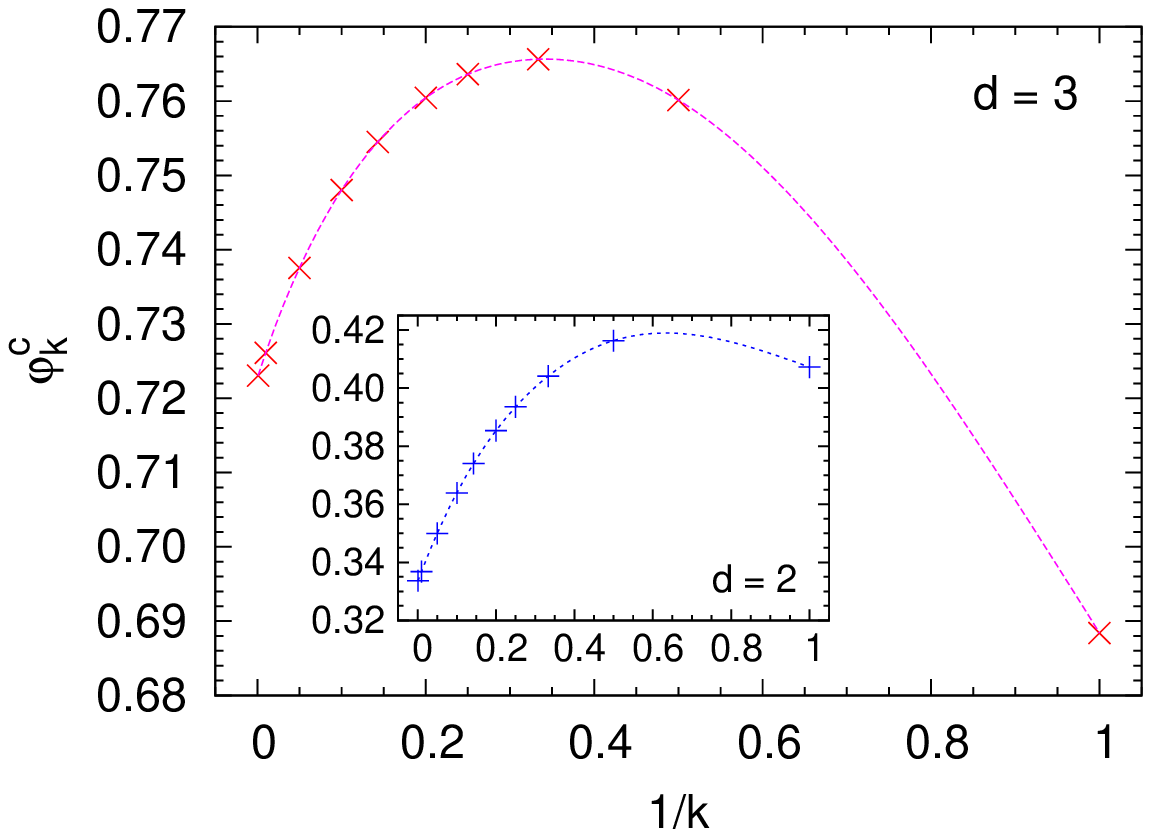} 
  \caption{ \label{fig:Results-I}
    (Color online)  Critical porosity $\varphi^\textrm{c}_k$ for Model I in 3D as a function of the reciprocal of the obstacle size, $1/k$.
    The dashed line is a guide to the eye drawn with cubic splines. Inset: the same for $d = 2$.}
 \end{center}
\end{figure}

The critical porosity for Model I (percolation of obstacles)
for several values of $k$ ranging from 1 to 10~000 is listed in table~\ref{tab:1}, and
its dependency on $1/k$ in two and three dimensions is depicted in figure~\ref{fig:Results-I}.
\begin{table}
   \caption{Critical porosity for selected values of $k$. The values for $k\to\infty$ come from extrapolation
           and are expected to be the same for $d=2$.
           The data for Model II and $k >20$ are not available.
           \label{tab:1}
          }
\begin{indented}
\item[]\begin{tabular}{rllll}
  \br
$k$  & \multicolumn{2}{c}{Model I} & \multicolumn{2}{c}{Model II}\\
  & \multicolumn{1}{c}{$d=2$}  & \multicolumn{1}{c}{$d=3$} & \multicolumn{1}{c}{$d=2$}  & \multicolumn{1}{c}{$d=3$}\\
    \mr
 1        & 0.40726(1)  & 0.68840(1)  & 0.5926(2)  & 0.3117(1) \\
 2        & 0.41635(2)  & 0.76013(2)  & 0.4868(1)  & 0.1687(1) \\
 3        & 0.40414(2)  & 0.76564(1)  & 0.4417(1)  & 0.1195(1) \\
 4        & 0.39352(1)  & 0.76362(1)  & 0.4172(2)  & 0.0961(1) \\
 5        & 0.38533(2)  & 0.76044(2)  & 0.4015(2)  & 0.0829(2) \\
 7        & 0.37403(1)  & 0.75450(1)  & 0.3829(3)  & 0.0685(2) \\
 10       & 0.36391(2)  & 0.74803(1)  & 0.3683(5)  & 0.0585(3) \\
 20       & 0.34994(2)  & 0.73754(2)  & 0.351(1)   & 0.048(1) \\
 $100$   & 0.33682(2)  & 0.72611(1)  &  \multicolumn{1}{c}{---}       &   \multicolumn{1}{c}{---}      \\
 $1~000$   & 0.33361(1)  & 0.72306(2)  &  \multicolumn{1}{c}{---}       &   \multicolumn{1}{c}{---}      \\
 $10~000$   & 0.33326(1)  & 0.72277(2)  &  \multicolumn{1}{c}{---}       &   \multicolumn{1}{c}{---}      \\
 $\infty$ & 0.33325(2)  & 0.72273(2)  & 0.334(1)   & 0.036(1)  \\
    \br
  \end{tabular}
 \end{indented}
\end{table}
A striking feature of these results is a nonmonotonic dependency of $\varphi^\textrm{c}_k$ on $k$.
This effect is particularly strong in three dimensions---in this case the maximum value of the critical porosity is located
at $k=3$ and exceeds its value for the standard site percolation ($k=1$) by $\approx 10\%$.

Our results are in agreement with those obtained for $k=1$ (standard site percolation) in references \cite{Jacobsen2014,Xu2014}.
Moreover, they allow to estimate the critical porosity in the limit of $k\to\infty$, which corresponds to the continuous
percolation of aligned squares or cubes. To this end we assumed that
$\varphi^\textrm{c}_k$  depends linearly on $1/k$ for sufficiently large $k$
and hence the value of $\varphi^\textrm{c}_\infty = \lim_{k\to\infty}\varphi^\textrm{c}_k$
can be estimated through linear extrapolation.
The critical surface/volume $\Phi_\textrm{c}$ of the continuous percolation is then related to
the critical porosity through $\Phi_\textrm{c} = 1 - \varphi^\textrm{c}_\infty$.
Our result for $d=2$, $\Phi_\textrm{c} = 0.66675(2)$, agrees with the value $0.66674349(3)$
obtained by Mertens and Moore \cite{Mertens2012}.
For the case $d=3$ we obtained $\Phi_\textrm{c} = 0.27727(2)$,
which is an improvement over the value 0.2773(2) reported by Baker et al.~\cite{Baker2002}.

\subsubsection{Model II \label{sub:Model-II}}

\begin{figure}
 \begin{center}
  \includegraphics[width=0.95\columnwidth]{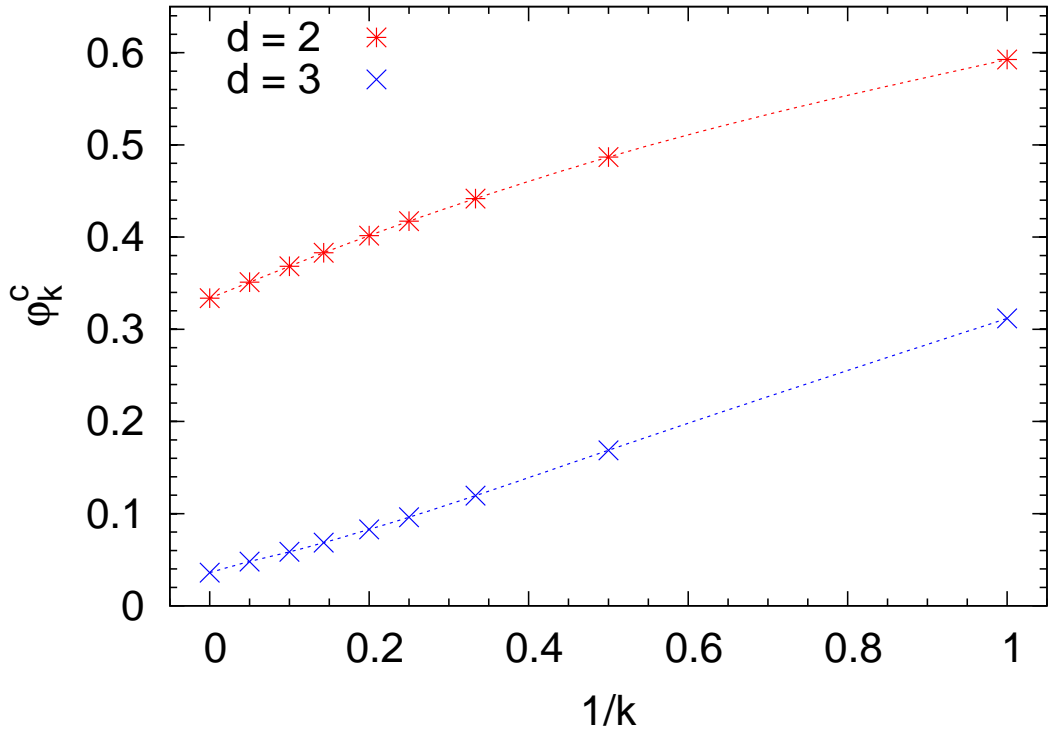}  
   \caption{ \label{fig:Results-II}
      (Color online) Critical porosity $\varphi^\textrm{c}_k$ for $d=2,3$ as a function of the reciprocal of
      the obstacle size, $1/k$, for Model II (percolation of voids).
      The values at $1/k=0$  come from extrapolation.
      The dashed lines are guides to the eye drawn with cubic splines.
}
 \end{center}
\end{figure}

\begin{figure}
 \begin{center}
  \includegraphics[width=0.5\columnwidth]{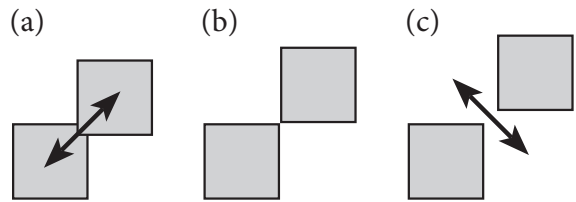} 
   \caption{ \label{fig:abc}
(a) Overlapping squares contribute to percolation of obstacles. (b) Two squares touching only at the corners
block percolation of both phases.  (c) Two disjoint squares facilitate percolation of the void phase. }
 \end{center}
\end{figure}

The critical porosity obtained for Model II (percolation of voids)
is listed in table~\ref{tab:1}, and
figure~\ref{fig:Results-II}
shows that its dependency  on $1/k$  is almost linear both for $d=2$ and $d=3$.
We verified that our results for $k=1$ are in agreement
with those reported recently in \cite{Jacobsen2014,Xu2014}.
As for the limit of $k\to\infty$, which corresponds to the continuum percolation of voids, this problem
is much harder than the continuous percolation of obstacles and
most of the research has been concentrated on overlapping discs or spheres.
Our results suggest that in this limit the critical porosities for Models I and II are the same if $d=2$.
This can be justified as follows. Consider two obstacles that are close to each other.
As shown schematically in figure~\ref{fig:abc},
if they  overlap or share a side (2D) or face (3D), they facilitate  the percolation of obstacles (Model I). If they are disjoint,
they facilitate percolation of the void phase (Model II). However, if they touch only at a corner (2D) or an edge (3D),
they block the percolation of both phases. The latter case is very common if $k$ is small,
but the probability of two squares sharing only a corner or two cubes sharing only
an edge quickly decreases as $k\to\infty$.
Thus, for continuous percolation the situation depicted in figure~\ref{fig:abc}b practically
does not occur and its impact is negligible. This implies that either phase A or B percolates regardless of the porosty.
As the percolation in one direction of a 2D system precludes the percolation of the other phase in the perpendicular direction,
and since the system is isotropic, the continuous percolation thresholds of aligned squares
of the obstacles (Model I) and of the void space around them (Model II) must be the same.
Consequently, even though our direct estimation of $\varphi^\textrm{c}_\infty$ in Model II yields $0.334(1)$,
one can safely use its more accurate approximation for Model~I, $0.33325651(3)$  \cite{Mertens2012}.

For $d=3$ the critical porosity in Model II is always smaller than  in Model I, which means
that for any $k$, including the case $k\to \infty$,  there exists a range of porosities for which the system resembles a sponge:
both the solid and void phases percolate. Moreover, the value of $\varphi^\textrm{c}_k$ for $k\to \infty$ turns out to be close to
the critical porosity 0.0317 of the void space around overlapping spheres \cite{Priour2014} (see also \cite{Marck1996}).
As expected, for $d=2$ we found the opposite behavior:
it is impossible to have a two-dimensional, infinite system with percolation of both phases.
Instead, for any finite $k$ there exists a range of porosities for which neither phase percolates.
The width of this range shrinks to zero as $k\to\infty$.
Note that our result for $k\to\infty$, $\varphi^\textrm{c}_\infty \approx 0.33324$,
is much larger than 0.159 reported by van der Marck \cite{Marck1996}
for continuous percolation of the void phase around disks. Such a large discrepancy cannot be explained
by the difference in the shape of obstacles. As Marck's result is
also smaller than the critical porosity of disks,
$\varphi^\textrm{c}_\infty \approx 1 - 0.67634831 \approx 0.32$ \cite{Mertens2012},
his result is probably erroneous.


\subsection{Discrete excluded volume approximation for Model I\label{sec:eva}}

According to the excluded volume theory \cite{Balberg1984,Balberg1987}, the continuum percolation threshold
for overlapping obstacles can be approximated with
\begin{equation}
 \label{eq:eva}
   n_\mathrm{c} V_\mathrm{ex} \approx  B,
\end{equation}
where $n_\mathrm{c} = \eta_\mathrm{c}/v$ is the critical number density of objects, $V_\mathrm{ex}$ is their
excluded area or volume, and $B$ is the average number of overlaps
per obstacle at criticality.  In continuous percolation $V_\mathrm{ex}$  is
defined as the area (volume) around an object into which
the center of another similar object is not allowed to enter
if overlapping of the two objects is to be avoided. Consequently,  $V_\mathrm{ex}$  can be interpreted as
a measure of the rate at which consecutive obstacles, which are free to overlap,
 form new connections as they are being placed in the system.
Therefore, for discrete models we define $V_\mathrm{ex}$ as the number of obstacle configurations which make a connection
with a given obstacle. In other words, for discrete models $V_\mathrm{ex}$
should be regarded as a ``connectedness factor'' rather than the excluded volume.
For Model I this leads to $V_\mathrm{ex}(k)$ being the number of elementary lattice cells
in a square or cube of length $2k+1$ that are not at its corners ($d=2$) or edges ($d=3$).
A general formula for $V_\mathrm{ex}(k)$  in Model I reads
\begin{equation}
  \label{eq:Vex-any-d}
    V_\mathrm{ex}(k) = (2k-1)^{d-1}(2k + 2d -1).
\end{equation}

Combining equations (\ref{eq:eta}) and (\ref{eq:eva}) one arrives at a general relation for the critical porosity
\begin{equation}
  \label{eq:approx-general}
      \varphi^\mathrm{c} \approx  \exp\left( -B\frac{v}{V_\mathrm{ex}} \right).
\end{equation}
For Model I it reduces to
\begin{equation}
  \label{eq:approx-I}
      \varphi^\mathrm{c}_k \approx
          \exp\left[
                       \displaystyle
                       -B_d\frac{k^d}{(2k-1)^{d-1}(2k + 2d -1)}
               \right],
\end{equation}
where $B_d$  is the $B$ factor for the space dimension $d$.

The ratio $V_\mathrm{ex}/v$ is  nonmonotonic in $k$,
with the maximum at a non-integer
\begin{equation}
 \label{eq:kmax}
  k^\mathrm{max}=d - 1/2.
\end{equation}
This corresponds to the nonmonotonic dependency
of the critical porosity on the obstacle size seen in figure \ref{fig:Results-I} for $d=2,3$.

Parameters $B_d$ in (\ref{eq:approx-I}) can be treated as fitting parameters to the simulation data or can be obtained
independently using other methods. In the former case we found that  $B_2 = 4.49$, $B_3 = 2.67$ and
$\varphi^\mathrm{c}_k $ determined from equation (\ref{eq:approx-I}) agree with the simulation results within 2 significant
digits (the absolute error $<  0.01$) for all $k$. In the later case one can use the values for the
continuous percolation: $B_2 = 4.3953711(5)$ and $B_3 = 2.5978(5)$.
They can be obtained by inserting the most accurate values of the critical porosity into
\begin{equation}
  B_d = -2^d\ln \varphi^\mathrm{c}_\infty ,
\end{equation}
which follows immediately from (\ref{eq:approx-general}) if we assume this relation to be exact. These values agree with
$B_2 = 4.39(1)$ and $B_3 = 2.59(1)$ reported in \cite{Baker2002}.

\begin{figure}
 \begin{center}
  \includegraphics[width=0.9\columnwidth]{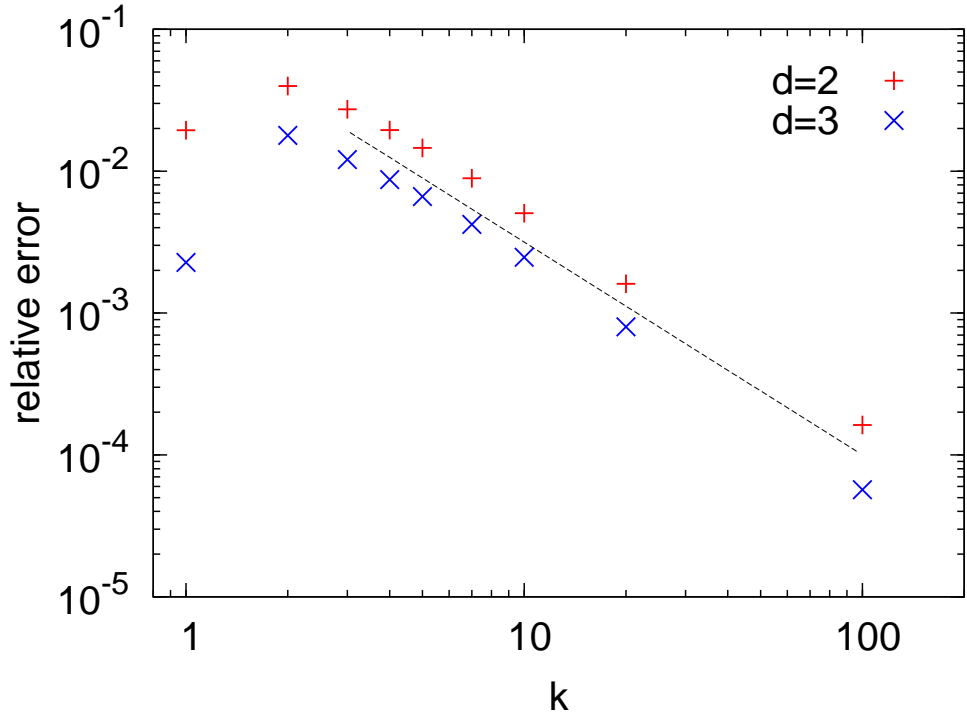}
   \caption{ \label{fig:eva}
   (Color online) The relative error
   of the critical porosity, $\varphi^\mathrm{c}_k$, approximated with equation (\protect\ref{eq:approx-I}),
   as a function of the object size $k$, for $d=2,3$ (Model I).
   The simulation results (table \protect\ref{tab:1}) were assumed to be the true values and $B_d$
   was taken from the continuous percolation. The dashed line is a guide to the eye with the slope $-3/2$.
   }
 \end{center}
\end{figure}

The error introduced by (\ref{eq:approx-I}) can be estimated using the simulation results as the reference values. As
shown in figure \ref{fig:eva},
even with the values of $B_d$ taken from the continuous percolation,
the accuracy of  (\ref{eq:approx-I}) is remarkable.
As expected, the quality of (\ref{eq:approx-I}) improves as $k$ goes to infinity. It is rather surprising
that the correction to (\ref{eq:approx-I}) scales in the same way, as $k^{-3/2}$, for both $d=2$ and $3$.
 It is also  worth noticing that
for $d=3$ the simple theory presented in this section allows to predict the critical point of site percolation
($k=1$) with the accuracy of $\approx 0.2\%$ using only the information available for the continuous percolation.

Encouraged by the success of the theory for $d=2,3$, one might ask whether it can be  applied in higher dimensions.
Since  $V_\mathrm{ex} = 2^d$
for continuous percolation   and $V_\mathrm{ex} = 2d+1$ for site percolation,
equation  (\ref{eq:approx-general}) suggests a general relation
between the critical porosity in the continuous ($k\to\infty$) and site ($k=1$) percolation on regular
lattices of arbitrary dimension $d$,
\begin{equation}
  \label{eq:phi-phi}
  \varphi^\mathrm{c}_1 \approx (\varphi^\mathrm{c}_\infty)^{2^d/2d+1}.
\end{equation}
This equation correctly predicts that
$\varphi^\mathrm{c}_1 < \varphi^\mathrm{c}_\infty$ for space dimensions 3 and higher.
Moreover, using the available results for the high-dimensional site percolation
\cite{Grassberger2003} and continuous percolation of oriented hypercubes \cite{Torquato2012b},
it is straightforward to verify  that for dimensions $d\le 5$ the
values of the site percolation threshold $1-\varphi^\mathrm{c}_1$ predicted from (\ref{eq:phi-phi}) deviate from the correct
ones by less than 3.5\%, whereas  for $6 \le d \le 11$ the accuracy varies erratically between 8 and 15\%.
One possible explanation is as follows.
One might expect that the quality of our approach is very good for large $k$
and worsens as $k$ goes down to 1,
especially for  $k \le k^\mathrm{max}$, i.e. behind the maximum value of $\varphi^\mathrm{c}(k)$.
Since $k^\mathrm{max}$ diverges with $d$,
the quality of (\ref{eq:phi-phi}), which involves the information for  $k=1$,
can be expected to decline for large $d$.
It is also possible that the smaller accuracy of  (\ref{eq:phi-phi}) for $d\ge6$ is related to 6
being the critical dimension for standard percolation \cite{Essam80}.


\subsection{Relation to the site percolation with complex neighborhood\label{sec:complex-neighborhood}}

\begin{figure}
 \begin{center}
  \includegraphics[width=0.75\columnwidth]{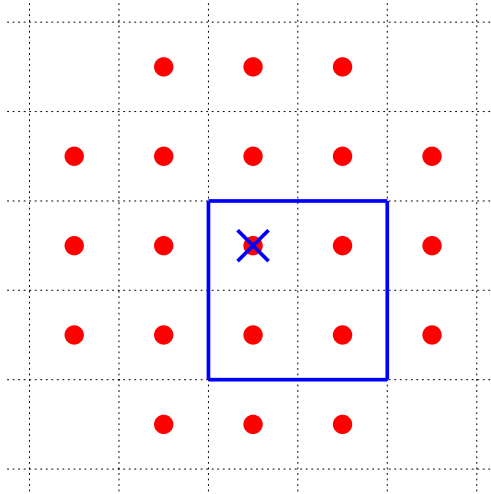}  
   \caption{ \label{fig:neigbourhood} (Color online) The relation between percolation of squares of size $k\times k$
and the standard site percolation with complex neighborhood.
See the text for detailed explanation.  }
 \end{center}
\end{figure}

Percolation of obstacles of linear size $k$ can be mapped on the standard site percolation with complex neighborhood problem.
Consider an example visualized in figure~\ref{fig:neigbourhood},
which shows the simplest nontrivial case $d=2$, $k=2$.
The position of each obstacle on the lattice can be identified by the coordinates of one of the elementary cells making up the obstacle,
e.g., the one occupying the top leftmost corner. In figure~\ref{fig:neigbourhood}
this characteristic elementary cell for the $2\times2$ obstacle drawn
with a thick solid line is marked by a cross. Another obstacle is connected to the reference one if and only if
its characteristic elementary cell is located in one of the elementary cells marked by circles in figure~\ref{fig:neigbourhood}
(which form the ``excluded volume'' $V_\mathrm{ex}$ for the reference object).
Thus, the connectedness  of a group of squares of linear size $k=2$ is equivalent to that of
elementary squares of linear size 1 with complex neighborhood marked with circles.
This observation can be generalized to arbitrary space dimension $d$, arbitrary value of $k$ and even arbitrary obstacle shape.

While the connectedness in the two models is equivalent, the porosities in each model are different---in
the site percolation with complex neighborhood
each obstacle occupies only one elementary cell. The porosity in the equivalent problem of site percolation, denoted henceforth by $\pi$,
can be calculated from (\ref{eq:<phi>}) with $v = 1$. This immediately leads to
\begin{equation}
  \label{eq:connection}
   \varphi^\textrm{c}_k = (\pi^\textrm{c}_k)^{k^d},
\end{equation}
where $\pi^\textrm{c}_k$ is the critical porosity in the equivalent site percolation with complex neighborhood problem.

The site percolation with complex neighborhood corresponding to the case illustrated in figure~\ref{fig:neigbourhood}
($d=2$, $k=2$) was investigated by Majewski and Malarz \cite{Majewski2007},
who found $1 - \pi^\textrm{c}_2 \approx 0.196$.
This is consistent with our result $1 - \pi^\textrm{c}_2 = 0.196724(5)$ that follows from (\ref{eq:connection}) and the
value of  $\varphi^\textrm{c}_2 = 0.41635(2)$ listed in table~\ref{tab:1}.


\section{Conclusions and outlook\label{sec:conclusions}}

The model of overlapping hypercubes, besides its applications in porous media modelling,
allows one to investigate the transition between discrete and continuous percolation.
We developed
a simple phenomenological theory, an extension of the excluded volume approximation to discrete systems,
which describes this transition with a remarkable accuracy. We verified this theory numerically in dimensions
2 and 3 and gave evidence that it is likely to remain valid in higher dimensions.
We were also able to improve the accuracy of the determination of the continuous percolation threshold
of aligned, overlapping cubes in three dimensions as well as to provide the values of continuous percolation thresholds
of the void space around aligned squares and cubes in dimensions two and three.
Moreover, we showed that a direct relation  between the model of overlapping
objects of arbitrary shape and the standard site percolation
with complex neighborhood can be used to improve the accuracy of the percolation
threshold determination in a class of percolation models.

We found that the percolation threshold in a model of aligned, overlapping squares or cubes
randomly distributed on a regular lattice is a nonmonotonic function of their size.
This is an unexpected behavior, potentially affecting porous media modelling.
Percolation of the void space around squares or cubes turns out to behave in a more regular way,
as in this case the critical porosity can be approximated by a function
linear in the reciprocal of the obstacle linear size.

Universality of the results presented here requires further studies,
especially on obstacles of needle-like shapes, an idea that proved  illuminating
for the development of the excluded volume theory
 \cite{Balberg1987,Provatas2000,Asikainen2000,Balberg1984}.


\section*{References}


\begin{thebibliography}{10}

\bibitem{Torquato2002}
Salvatore Torquato.
\newblock {\em Random heterogeneous materials: microstructure and macroscopic
  properties}.
\newblock Springer, New York, 2002.

\bibitem{Weissberg63}
Harold~L. Weissberg.
\newblock Effective diffusion coefficient in porous media.
\newblock {\em J. Appl. Phys.}, 34(9):2636--2339, 1963.

\bibitem{Maxwell}
J.~C. Maxwell.
\newblock {\em A Treatise on Electricity and Magnetism}, volume~I.
\newblock Clarendon Press, London, 2 edition, 1881.

\bibitem{Voutilainen2013}
Mikko Voutilainen, Paul Sardini, Marja Siitari-Kauppi, Pekka Kek+¥l+¥inen, Vesa
  Aho, Markko Myllys, and Jussi Timonen.
\newblock Diffusion of tracer in altered tonalite: {E}xperiments and
  simulations with heterogeneous distribution of porosity.
\newblock {\em Transport in Porous Media}, 96(2):319--336, 2013.

\bibitem{Zalc2003}
J.~M. Zalc, S. Reyes, and E. Iglesia.
\newblock Monte-{C}arlo simulations of surface and gas phase diffusion in
  complex porous structures.
\newblock {\em Chemical Engineering Science}, 58(20):4605 -- 4617, 2003.

\bibitem{Mace1991}
Olivier Mace and James Wei.
\newblock Diffusion in random particle models for hydrodemetalation catalysts.
\newblock {\em Industrial \& Engineering Chemistry Research}, 30(5):909--918,
  1991.

\bibitem{Baker2002}
Don~R. Baker, Gerald Paul, Sameet Sreenivasan, and H.~Eugene Stanley.
\newblock Continuum percolation threshold for interpenetrating squares and
  cubes.
\newblock {\em Phys. Rev. E}, 66:046136, Oct 2002.

\bibitem{Torquato2012a}
S.~Torquato.
\newblock Effect of dimensionality on the continuum percolation of overlapping
  hyperspheres and hypercubes.
\newblock {\em J. Chem. Phys.}, 136:054106, Oct 2012.

\bibitem{Torquato2012b}
S.~Torquato and Y.~Jiao.
\newblock Effect of dimensionality on the continuum percolation of overlapping
  hyperspheres and hypercubes. {II}. {S}imulation results and analyses.
\newblock {\em The Journal of Chemical Physics}, 137(7):074106, 2012.

\bibitem{Garboczi1991}
E.~J. Garboczi, M.~F. Thorpe, M.~S. DeVries, and A.~R. Day.
\newblock Universal conductivity curve for a plane containing random holes.
\newblock {\em Phys. Rev. A}, 43:6473--6482, Jun 1991.

\bibitem{Asikainen2000}
J.~Asikainen and T.~Ala-Nissila.
\newblock Percolation and spatial correlations in a two-dimensional continuum
  deposition model.
\newblock {\em Phys. Rev. E}, 61:5002--5008, May 2000.

\bibitem{Lorenz2001}
Christian~D. Lorenz and Robert~M. Ziff.
\newblock Precise determination of the critical percolation threshold for the
  three-dimensional ``{S}wiss cheese'' model using a growth algorithm.
\newblock {\em J. Chem. Phys.}, 114:3659, Feb 2001.

\bibitem{Priour2014}
D.~J. Priour.
\newblock Percolation through voids around overlapping spheres: {A} dynamically
  based finite-size scaling analysis.
\newblock {\em Phys. Rev. E}, 89:012148, Jan 2014.

\bibitem{Marck1996}
S.~C. van~der Marck.
\newblock Network approach to void percolation in a pack of unequal spheres.
\newblock {\em Phys. Rev. Lett.}, 77:1785--1788, Aug 1996.

\bibitem{Okazaki1996}
A.~Okazaki, K.~Maruyama, K.~Okumura, Y.~Hasegawa, and S.~Miyazima.
\newblock Critical exponents of a two-dimensional continuum percolation system.
\newblock {\em Phys. Rev. E}, 54:3389--3392, Oct 1996.

\bibitem{Balberg1987}
I.~Balberg.
\newblock Recent developments in continuum percolation.
\newblock {\em Philosophical Magazine Part B}, 56(6):991--1003, 1987.

\bibitem{Provatas2000}
N~Provatas, M~Haataja, J~Asikainen, S~Majaniemi, M~Alava, and T~Ala-Nissila.
\newblock Fiber deposition models in two and three spatial dimensions.
\newblock {\em Colloids and Surfaces A: Physicochemical and Engineering
  Aspects}, 165:209 -- 229, 2000.

\bibitem{Mertens2012}
Stephan Mertens and Cristopher Moore.
\newblock Continuum percolation thresholds in two dimensions.
\newblock {\em Phys. Rev. E}, 86:061109, Dec 2012.

\bibitem{Xia1988}
W.~Xia and M.~F. Thorpe.
\newblock Percolation properties of random ellipses.
\newblock {\em Phys. Rev. A}, 38:2650--2656, Sep 1988.

\bibitem{Yi2006}
Y.~B. Yi.
\newblock Void percolation and conduction of overlapping ellipsoids.
\newblock {\em Phys. Rev. E}, 74:031112, Sep 2006.

\bibitem{Kohout2004}
M.~Kohout, A.P. Collier, and F.~{\v{S}}t{\v{e}}p{\'a}nek.
\newblock Effective thermal conductivity of wet particle assemblies.
\newblock {\em International Journal of Heat and Mass Transfer}, 47(25):5565 --
  5574, 2004.

\bibitem{Matyka2013}
Maciej Matyka, Zbigniew Koza, Jaros\l{}aw Go\l{}embiewski, Marcin Kostur, and
  Micha\l{} Januszewski.
\newblock Anisotropy of flow in stochastically generated porous media.
\newblock {\em Phys. Rev. E}, 88:023018, Aug 2013.

\bibitem{Koponen1996}
A.~Koponen, M.~Kataja, and J.~Timonen.
\newblock Tortuous flow in porous media.
\newblock {\em Phys. Rev. E}, 54:406--410, Jul 1996.

\bibitem{Koponen1997}
A.~Koponen, M.~Kataja, and J.~Timonen.
\newblock Permeability and effective porosity of porous media.
\newblock {\em Phys. Rev. E}, 56:3319--3325, Sep 1997.

\bibitem{Matyka2008}
Maciej Matyka, Arzhang Khalili, and Zbigniew Koza.
\newblock Tortuosity-porosity relation in porous media flow.
\newblock {\em Phys. Rev. E}, 78:026306, Aug 2008.

\bibitem{Jiang2008}
Fangming Jiang and Antonio~C.M. Sousa.
\newblock Smoothed particle hydrodynamics modeling of transverse flow in
  randomly aligned fibrous porous media.
\newblock {\em Transport in Porous Media}, 75(1):17--33, 2008.

\bibitem{Balberg1984}
I.~Balberg, C.~H. Anderson, S.~Alexander, and N.~Wagner.
\newblock Excluded volume and its relation to the onset of percolation.
\newblock {\em Phys. Rev. B}, 30:3933--3943, Oct 1984.

\bibitem{Stauffer1994}
D.~Stauffer and A.~Aharony.
\newblock {\em Introduction to Percolation Theory}.
\newblock Taylor and Francis, London, 2 edition, 1994.

\bibitem{Isichenko1992}
M.~B. Isichenko.
\newblock Percolation, statistical topography, and transport in random media.
\newblock {\em Rev. Mod. Phys.}, 64:961--1043, Oct 1992.

\bibitem{Cornette2003}
V.~Cornette, A.J. Ramirez-Pastor, and F.~Nieto.
\newblock Percolation of polyatomic species on a square lattice.
\newblock {\em The European Physical Journal B -- Condensed Matter and Complex
  Systems}, 36:391--399, Dec 2003.

\bibitem{Xu2014}
Xiao Xu, Junfeng Wang, Jian-Ping Lv, and Youjin Deng.
\newblock Simultaneous analysis of three-dimensional percolation models.
\newblock {\em Frontiers of Physics}, 9(1):113--119, 2014.

\bibitem{Wang2013}
Junfeng Wang, Zongzheng Zhou, Wei Zhang, Timothy~M. Garoni, and Youjin Deng.
\newblock Bond and site percolation in three dimensions.
\newblock {\em Phys. Rev. E}, 87:052107, May 2013.

\bibitem{Jacobsen2014}
Jesper~Lykke Jacobsen.
\newblock High-precision percolation thresholds and {P}otts-model critical
  manifolds from graph polynomials.
\newblock {\em Journal of Physics A: Mathematical and Theoretical},
  47(13):135001, 2014.


\bibitem{Grassberger2003}
Peter Grassberger.
\newblock Critical percolation in high dimensions.
\newblock {\em Phys. Rev. E}, 67:036101, Mar 2003.

\bibitem{Essam80}
J.~W. Essam.
\newblock Percolation theory.
\newblock {\em Reports on Progress in Physics}, 43(7):833, 1980.

\bibitem{Majewski2007}
M.~{Majewski} and K.~{Malarz}.
\newblock Square lattice site percolation thresholds for complex
  neighbourhoods.
\newblock {\em Acta Physica Polonica B}, 38:2191, June 2007.

\end{thebibliography}

\end{document}